# Local spectroscopy of moiré-induced electronic structure in gate-tunable twisted bilayer graphene

Dillon Wong (黃家和),[1,2] Yang Wang (汪洋),[1,2] Jeil Jung (정재일),[3,4,5] Sergio Pezzini,[1,8] Ashley M. DaSilva,[4] Hsin-Zon Tsai,[1,2] Han Sae Jung,[1,2,10] Ramin Khajeh,[1] Youngkyou Kim (김영규),[1,11] Juwon Lee (이주원),[1,2] Salman Kahn (成吉思汗),[1,2] Sajjad Tollabimazraehno,[1,9] Haider Rasool,[1] Kenji Watanabe,[7] Takashi Taniguchi,[7] Alex Zettl,[1,2] Shaffique Adam,[5,6] Allan H. MacDonald,[4] and Michael F. Crommie[1,2,*]

[1] Department of Physics, University of California at Berkeley, Berkeley CA, 94720, United States

[2] Materials Science Division, Lawrence Berkeley National Laboratory, Berkeley CA, 94720, United States

[3] Department of Physics, University of Seoul, Seoul 130-743, Korea

[4] Department of Physics, University of Texas at Austin, Austin TX, 78712, United States

[5] Physics Department, National University of Singapore, 117551, Singapore

[6] Yale-NUS College, 6 College Avenue East, 138614, Singapore

[7] National Institute for Materials Science, Tsukuba, Ibaraki 305-0044, Japan

[8] Dipartimento di Fisica, Università degli studi di Pavia, I-27100 Pavia, Italy

[9] Johannes Kepler University Linz, Altenbergerstraße 69, 4040 Linz, Austria

[10] Department of Chemistry, University of California at Berkeley, Berkeley CA, 94720, United States

[11] Department of Chemical and Biomolecular Engineering, University of California at Berkeley, Berkeley CA, 94720, United States

*crommie@berkeley.edu

## Abstract

Twisted bilayer graphene (tBLG) forms a quasicrystal whose structural and electronic properties depend on the angle of rotation between its layers. Here we present a scanning tunneling microscopy study of gate-tunable tBLG devices supported by atomically-smooth and chemically inert hexagonal boron nitride (BN). The high quality of these tBLG devices allows identification of coexisting moiré patterns and moiré super-superlattices produced by graphene-graphene and graphene-BN interlayer interactions. Furthermore, we examine additional tBLG spectroscopic features in the local density of states beyond the first van Hove singularity. Our experimental data is explained by a theory of moiré bands that incorporates *ab initio* calculations and confirms the strongly non-perturbative character of tBLG interlayer coupling in the small twist-angle regime.

Van der Waals heterostructures [1] built from rotated layered materials exhibit novel electronic structure, including van Hove singularities (VHS) [2], massive Dirac fermions [3], superlattice Dirac points [4], and the Hofstadter butterfly [5,6]. Twisted bilayer graphene (tBLG) is a model system for such studies since it is composed of two identical stacked sheets of graphene oriented with a relative twist angle between them. Interactions between the two graphene layers result in a moiré pattern with a superlattice period that scales inversely with twist angle [7-10] and produce dramatic modifications to the graphene electronic properties [11-20]. For example, overlap between the two layers' Dirac cones leads to an avoided crossing and a resulting VHS peak in the density of states (DOS) [2,9,21]. Although the formation of the VHS has previously been reported for tBLG samples, the substrates used in those studies (silicon carbide, graphite, and metals) influence the tBLG electronic structure and thus complicate its interpretation [2,9,22,23]. This has led to discrepancies between measurements performed on different tBLG samples [2,9,23]. Electronic structure measurements of tBLG on less interacting substrates, such as hexagonal boron nitride (BN) [24-27], are therefore necessary to understand the intrinsic properties of tBLG.

Here we report a scanning tunneling microscopy (STM) study of electrostatically gated tBLG supported by a BN substrate (tBLG/BN). The high quality of our tBLG/BN devices enables us to visualize new tBLG phenomena on multiple length scales, from atomic lattices to moiré super-superlattices. Although coexisting moiré patterns have previously been observed in graphitic multilayers [9,10], we demonstrate here that moiré patterns produced by graphene-graphene interlayer interactions can coincide with moiré patterns that arise from graphene-BN interactions to produce a new super-superlattice structure. Our scanning tunneling spectroscopy (STS) measurements of tBLG/BN yield new insight into how the intrinsic electronic behavior of tBLG is affected by moiré-induced interactions. For example, we observe an additional dip feature in the DOS at an energy beyond the first VHS. By comparing the moiré wavelength dependence of this new feature's energy to calculations incorporating *ab initio* results, we find that it arises from a partial gap opening between the second and third moiré bands due to strong non-perturbative interlayer coupling in the small twist-angle regime. These measurements on our high quality tBLG/BN samples provide a first look into the atomic-scale behavior of electrostatically gated tBLG, an important device configuration for future applications [28,29].

Our measurements were performed using an Omicron ultra-high vacuum (UHV) STM operating at temperature T = 5 K with electrochemically etched PtIr tips. The tips were calibrated against the surface state of an Au(111) crystal before performing all measurements [30]. Differential conductance (d$I$/d$V$) was measured by lock-in detection of the a.c. tunnel current modulated by a 6-8 meV (rms), 400-700 Hz signal added to the sample bias ($V_s$). Each tBLG device was fabricated by growing two monolayers of graphene via chemical vapor deposition (CVD) [31] and then sequentially transferring the layers onto a BN flake peeled onto an SiO$_2$/Si wafer (see Fig. 1(a) for a schematic of a typical graphene device; see Ref. [32] for more details about the growth and transfer processes). The samples were annealed in UHV at 400°C for several hours to clean them before loading them into the STM.

The CVD-grown graphene used here was multi-crystalline with grain sizes on the order of several microns, and so the stacking of the two graphene layers naturally resulted in a multitude of random twist angles $\theta_{g-g}$ (angle between the two graphene layers) and $\theta_{g-BN}$ (angle between the bottom graphene layer and BN) across the entire sample. This is confirmed by the STM topographic images of Figs. 1(b)-1(d) which display different areas exhibiting different moiré wavelengths from different spots on the same sample. Figures 1(b) and 1(c) show graphene-graphene moiré patterns coexisting with graphene-BN moiré patterns, with the atomic lattice of the top graphene layer also visible (the atomic-scale structure can be more clearly seen in a higher magnification image in the Supplementary Material). The presence of these moiré patterns indicates that the interfaces between the graphene and BN layers are atomically sharp and free from contamination. The origins of the different moiré patterns are easily distinguished since the depressions in the graphene-graphene moiré pattern form a honeycomb-like appearance, whereas the depressions in the graphene-BN moiré pattern resemble a triangular lattice. Figure 1(b) exhibits a graphene-graphene moiré wavelength $\lambda_{g-g} = 1.4$ nm and a graphene-BN moiré wavelength $\lambda_{g-BN} = 12$ nm, while Fig. 1(c) has $\lambda_{g-g} = 1.2$ nm and $\lambda_{g-BN} = 3.3$ nm.

In contrast to Figs. 1(b) and 1(c) where $\lambda_{g-BN}$ is much larger than $\lambda_{g-g}$, Fig. 1(d) depicts a pair of coexisting moiré patterns where $\lambda_{g-BN} \approx \lambda_{g-g}$. When two moiré patterns are of similar size, they interfere to produce a moiré of the moirés. This new "super-superlattice" structure can be shown to have a wavelength given by

$$\lambda_S = \frac{\lambda_{g-BN}}{\sqrt{\left(\frac{\lambda_{g-BN}}{\lambda_{g-g}}\right)^2 + 1 - 2\left(\frac{\lambda_{g-BN}}{\lambda_{g-g}}\right)\cos(\Delta\theta)}}$$

where $\Delta\theta$ is the rotation angle between the graphene-graphene moiré and the graphene-BN moiré. We are able to extract $\lambda_{g-BN} = 4.7$ nm, $\lambda_{g-g} = 4$ nm, and $\Delta\theta = 18°$ through a Fourier transform of Fig. 1(d) (see the Supplementary Material). These values yield $\lambda_S = 12$ nm, consistent with direct measurement of the super-superlattice periodicity in Fig. 1(d).

In order to explore the intrinsic behavior of tBLG, we performed STM measurements on regions of our device with $\theta_{g-BN} > 15°$. Atomically-resolved topographic images of such regions are depicted in Fig. 2 and do not show any features arising from the underlying BN substrate. The twist angle $\theta_{g-g}$ for each region is obtained through the observed moiré period $\lambda_{g-g}$ with the relation [7,12]

$$\lambda_{g-g} = \frac{a}{2\sin\frac{\theta_{g-g}}{2}}$$

where $a = 2.46$ Å is the graphene lattice constant. We verify that the moiré patterns are created by the graphene-graphene interaction and not graphene-BN interaction by comparing the relative orientations of the top-layer graphene lattice and the moiré lattice. Figures 2(a), 2(b), and 2(c) display graphene-graphene moiré wavelengths 0.9 nm, 2.4 nm, and 6.1 nm, corresponding to $\theta_{g-g}$ twist angles 16°, 5.9°, and 2.3°, respectively. As the twist angle decreases, the local AA stacked regions (where top and bottom graphene sublattices are aligned) and AB stacked regions (where Bernal stacking occurs) become better defined (see Fig. 2(c)). Each AA region displays internal honeycomb-like structure, while the AB/BA regions display internal triangular-like lattice structures with opposite orientations (consistent with expected tBLG structure [13,33-36]).

The local electronic structure of tBLG/BN in the $\theta_{g-BN} > 15°$ regime was measured by performing STM d$I$/d$V$ spectroscopy in areas having different moiré wavelength $\lambda_{g-g}$ (see Supplementary Material for spectroscopy in areas with $\theta_{g-BN} < 15°$). Figure 3 shows d$I$/d$V$ spectra obtained in the AA regions of different moiré patterns (spectra obtained in AB/BA and bridge moiré regions are quite similar; see the Supplementary Material). All of the spectra in Fig. 3 were taken while holding the Dirac point close to (but not exactly at) the Fermi level via application of a gate voltage ($V_g$). The Dirac point feature is not clearly visible in these spectra because of a ~130 meV wide gap-like feature at the Fermi level that is caused by phonon-assisted inelastic tunneling (similar to the case for monolayer graphene [37]). Our spectra exhibit a peak and two dips (labeled by colored arrows) that disperse with moiré wavelength. The peak (black arrows) corresponds to the VHS previously reported by STS and angle-resolved photoemission spectroscopy [2,9,22], the 1st dip (red arrows) is a minima that follows the VHS, while the 2nd dip (green arrows) is a phenomenon not present in previous studies of tBLG on graphite or silicon carbide substrates [2,9]. The energy magnitudes of these features (relative to the Dirac point) decrease as the moiré wavelength increases, and each feature has a symmetric copy lying on the opposite side of the Dirac point.

Figure 4(a) shows a plot of the energy magnitudes of the three observed features (relative to the Dirac point) versus the moiré wavelength $\lambda_{g-g}$ (error bars are approximately the size of the symbols). The gate-tunability of our tBLG devices was crucial to maintain accuracy for this measurement since the measurement had to be performed under a condition of high carrier density (mid $10^{12}$ cm$^{-2}$) by setting the back-gate voltage $V_g$ to +60 V or -60 V. This allowed the Fermi level ($E_F$) to be positioned such that DOS($E_F$) is as high as possible, thus reducing the effect of tip-induced band-bending compared to the case where $E_F$ is positioned near the Dirac point [38] (which has low DOS) (see the Supplementary Material for a comparison between spectroscopic features measured at different carrier doping levels).

We carried out theoretical simulations and compared them to our experimental data in order to interpret the observed spectroscopic features. While the tBLG VHS is typically thought of as a saddle point generated by avoided crossing between rotated Dirac cones [2,9,39-41], it can more generally be thought of as the result of a collection of zone-folded Bloch bands arising from the periodicity associated with the moiré pattern [42]. In this picture the avoided crossing that gives rise to the VHS (i.e. the 1st dip) is the energy gap between the first and the second moiré Bloch bands, and higher order gaps are also expected. We modeled this behavior through an effective Hamiltonian formalism that allowed us to calculate the tBLG electronic band structure, total DOS, and local density of states (LDOS) for arbitrary bilayer twist angles [43] (see the Supplementary Material for LDOS simulations). This technique incorporates *ab initio* calculations performed in commensurate periodic structures as input to construct the effective Hamiltonian, and allows vertical relaxation of the interlayer separation (see Ref. [43] for details). This theory has no adjustable parameters other than the value of the unperturbed graphene Fermi velocity $v_F = 1.05 \times 10^6$ m/s (the Supplementary Material shows calculations for different values of $v_F$). A schematic plot of the momentum space structure of tBLG in a repeated zone scheme is shown in Fig. 4(b). The green line here represents the Wigner-Seitz cell of the moiré reciprocal lattice, while the red and black circles mark the Dirac points of the two graphene layers. Special high symmetry points in the Brillouin zone are labeled A, B, C, and D. The resulting moiré-induced band structure of tBLG for $\lambda_{g-g} = 7.6$ nm is plotted in Fig. 4(c) (band structure plots for different $\lambda_{g-g}$ are shown in the Supplementary Material). The corresponding integrated DOS is displayed in Fig. 4(d) and shows features that match our experimental results reasonably well (e.g. a peak and two dips, highlighted in semi-transparent color bars). A good match between theory and experiment is also seen for the energy magnitudes of the calculated spectroscopic features as a function of moiré wavelength (Fig. 4(a)), as well as for our energy-dependent d$I$/d$V$ maps (see the Supplementary Material).

The quantitative agreement between our theoretical calculations and experimental data enables us to identify the physical origin of the experimentally observed features. For example, the 1st dip seen in the d$I$/d$V$ spectroscopy can be identified as a partial gap opening between the first and second moiré bands (red highlighted bars in Figs. 4(c) and 4(d)), while the new 2nd dip feature is seen to correspond to the gap between the second and third moiré bands (green highlighted bars in Figs. 4(c) and 4(d)). Our study thus demonstrates the validity of the moiré bands picture of tBLG even for incommensurate orientations of the two graphene layers.

The moiré wavelength dependence of the spectroscopic features observed here also reveals the non-perturbative nature of the graphene interlayer interaction. As seen in Fig. 4(a), the energy of the VHS scales inversely with moiré wavelength for both theory and experiment. However, this is not the case for the 1st dip, where $1/\lambda_{g-g}$ scaling is not preserved for larger moiré periods (see the Supplementary Material). Inverse scaling with the moiré period, a signature of the perturbative regime, only occurs for large twist angles (small moiré wavelengths). For small twist angles (long moiré wavelengths) we find deviations from inverse scaling for the 1st dip, a signature of the non-perturbative interaction between the two graphene layers in this regime.

Due to this non-perturbative interaction, tBLG is quite different from the case of monolayer graphene supported by BN. The interaction between monolayer graphene and BN induces a simple periodic perturbation which gives rise to superlattice-induced mini-Dirac points in graphene [4,44-46]. Accurate parameterization of the moiré pattern site energy, local gap, and pseudogauge virtual strain terms has been presented for graphene on BN, but the coupling between two graphene layers in tBLG, on the other hand, cannot be reduced to effective periodic potentials within each independent layer [43]. The strong interlayer coupling in tBLG can be seen by examining the band structure plot of Fig. 4(c) more carefully. Two Dirac-cone-like features are seen at points C and D near the energy window of the 1st dip (red bar), but no mini-Dirac cones are seen in the energy window of the 2nd dip (green bar). Overall, the moiré band features of tBLG are in stark contrast to the case of monolayer graphene on BN [46].

In conclusion, we have fabricated gate-tunable tBLG devices using high quality BN substrates and demonstrated that graphene-graphene moiré patterns can exist simultaneously with graphene-BN moiré patterns. The existence of these coexisting moiré patterns may have important consequences for device applications involving stacked graphene and BN layers, and the emergence of graphene-BN super-superlattices may serve as an interesting platform for novel physical phenomena. The high quality of our tBLG/BN devices has enabled us to probe the intrinsic electronic properties of tBLG and to observe new spectroscopic features in the tBLG DOS. These features occur at energies beyond the first VHS and can be explained by a non-perturbative theory involving *ab initio* calculations. These results show that the strong coupling, low twist-angle regime in tBLG can be accessed using scanned probe techniques via high quality devices that allow electrostatic Fermi level tuning.

We thank Feng Wang for useful discussions. This research was supported by the Department of Energy under contract no. DE-AC02-05CH11231 (sp$^2$ program) (STM imaging and spectroscopy) and contract no. DE-FG02-ER45118 (development of effective Hamiltonian formalism), National Science Foundation grant DMR-1206512 (sample fabrication), the National Research Foundation of Singapore under its Fellowship program NRF-NRFF2012-01 (tBLG DOS and LDOS simulations), and the Elemental Strategy Initiative conducted by the MEXT Japan (BN synthesis). D.W. was supported by the Department of Defense (DoD) through the National Defense Science & Engineering Graduate Fellowship (NDSEG) Program, 32 CFR 168a. T.T. acknowledges support from a Grant-in-Aid for Scientific Research no. 262480621 and Innovative Areas "NanoInformatics" no. 25106006 from JSPS. D.W., Y.W., and J.J. contributed equally to this work.

**Figure Captions:**

**Figure 1. STM topographic images of tBLG/BN.** (**a**) Gated tBLG/BN field effect transistor. The tip bias (-$V_s$), back-gate voltage ($V_g$), and grounding scheme are shown. (**b-d**) STM topographic images show coexisting graphene-graphene and graphene-BN moiré patterns. Tunneling parameters: (**b**) $V_s$ = -0.2 V, I = 0.30 nA, $V_g$ = 0 V; (**c**) $V_s$ = -0.3 V, I = 0.30 nA, $V_g$ = +30 V; (**d**) $V_s$ = -3.0 V, I = 0.02 nA, $V_g$ = 0 V.

**Figure 2. STM topographic images of tBLG/BN for different graphene-graphene twist angles.** (**a-c**) STM topographic images of tBLG show both the top-layer atomic lattices and the graphene-graphene moiré patterns for decreasing twist angle $\theta_{g-g}$. No graphene-BN moiré patterns are visible in these images since $\theta_{g-BN} > 15°$. Tunneling parameters: (**a**) $V_s$ = -0.5 V, I = 0.1 nA, $V_g$ = 0 V; (**b**) $V_s$ = -0.5 V, I = 0.2 nA, $V_g$ = +30 V; (**c**) $V_s$ = -0.1 V, I = 0.1 nA, $V_g$ = -60 V.

**Figure 3. STM d$I$/d$V$ spectra of tBLG for different graphene-graphene moiré wavelengths.** dI/dV spectra reveal three features (VHS peak labeled by black arrow, 1st dip labeled by red arrow, and 2nd dip labeled by green arrow) that disperse with graphene-graphene moiré wavelength $\lambda_{g-g}$. Curves have been shifted vertically for clarity. Initial tunneling parameters: $V_s$ = 1 V, 0.2 nA $\leq$ I $\leq$ 0.4 nA, a.c. modulation 6 mV $\leq V_{rms} \leq$ 8 mV.

**Figure 4. Calculated tBLG electronic structure.** (**a**) Dependence of theoretical tBLG spectroscopic features on moiré wavelength $\lambda_{g-g}$ (dashed lines) compared to experiment (symbols). Experimental error bars are approximately the size of the symbols. (**b**) Schematic plot of the repeated-zone momentum

space structure of tBLG. The Wigner-Seitz cell (green line) of the moiré reciprocal lattice and the Dirac points of the two graphene layers (red and black circles) are shown. A – D mark high symmetry points in the moiré Brillouin zone. (**c**) Calculated moiré-induced band structure of tBLG for $\lambda_{g-g} = 7.6$ nm plotted along k-space lines connecting points A-B-C-D-A in **b**. Gray, red, and green highlighting locates the energies of the VHS peak, the 1st dip, and the 2nd dip, respectively. (**d**) Theoretical DOS of tBLG for $\lambda_{g-g} = 7.6$ nm calculated by integrating the band structure over the moiré Brillouin zone and then Gaussian broadening with a 15 meV width (the broadening represents the effects of temperature, finite quasiparticle lifetime, and a.c. wiggle voltage). The semi-transparent highlighting bars are located at the same energies as in **c** and match the three spectroscopic features observed in the experiment.

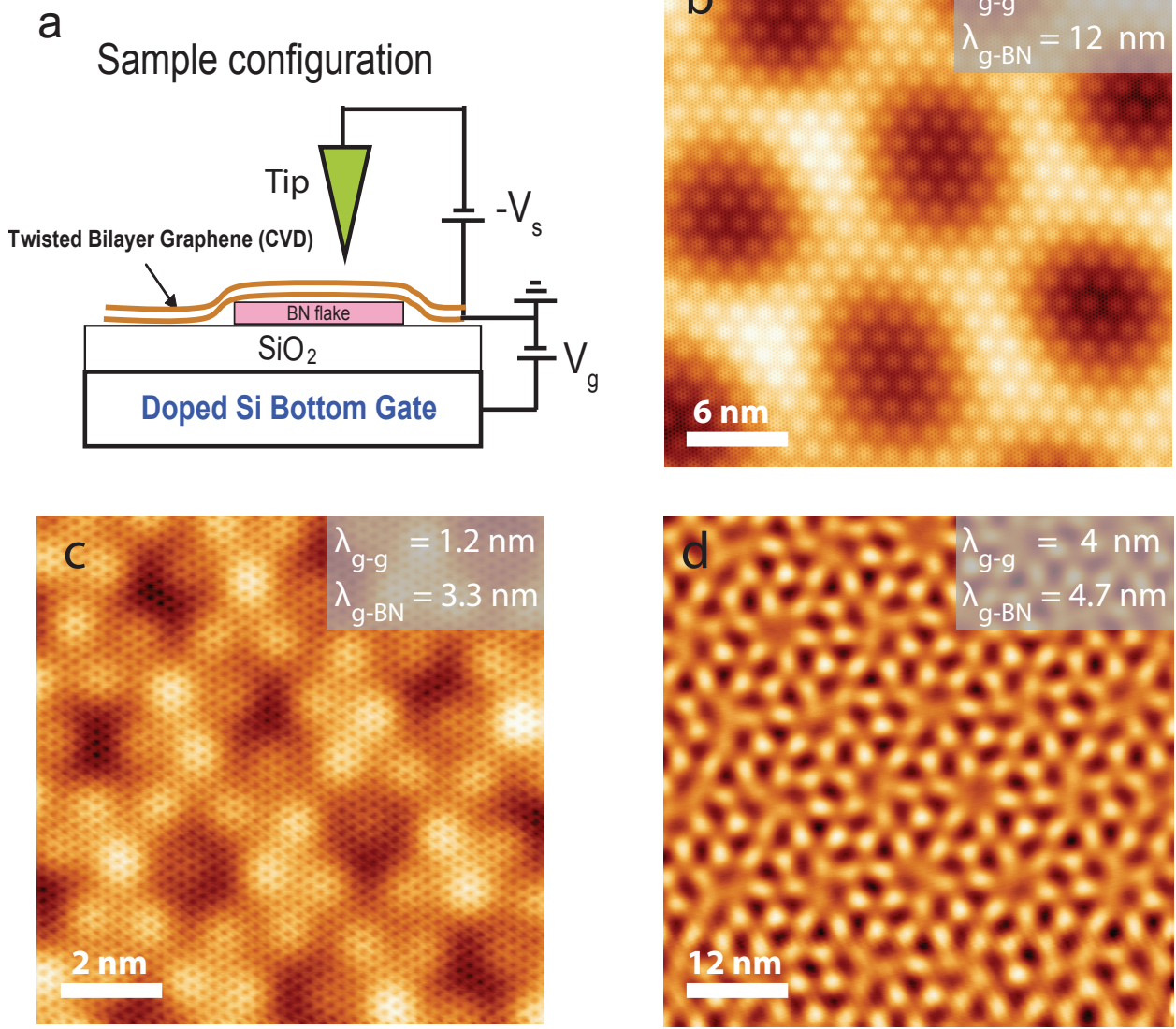

Figure 1

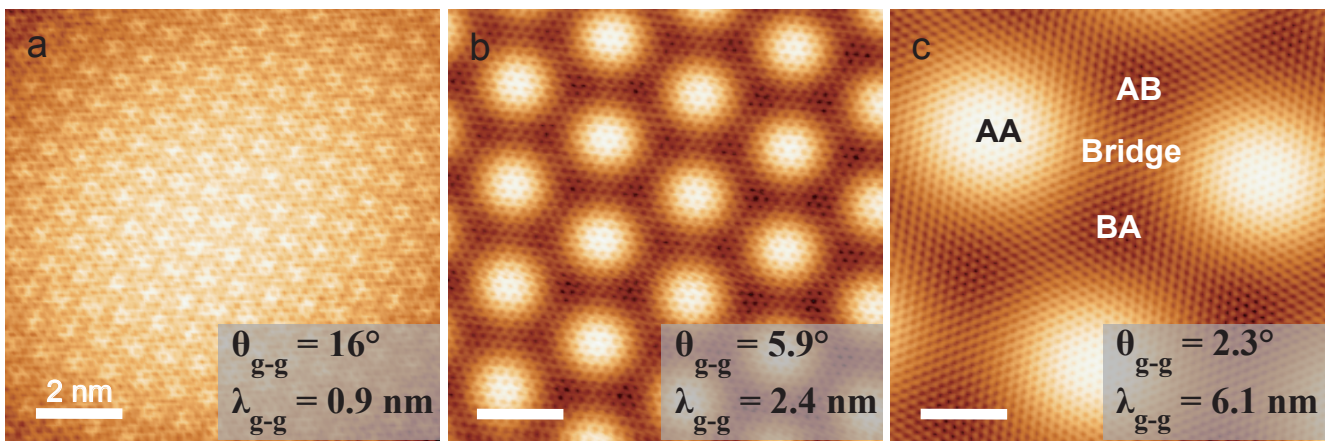

Figure 2

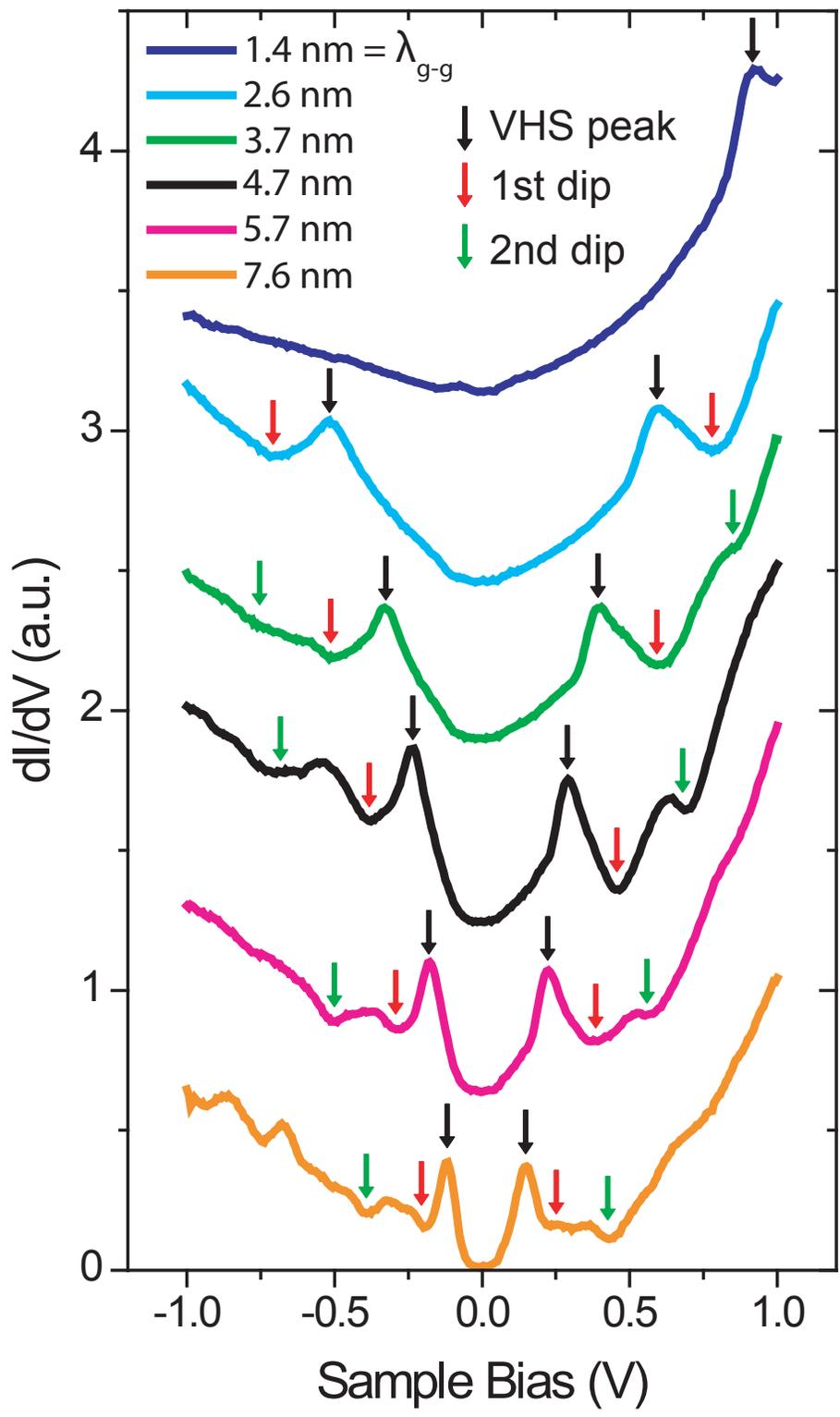

Figure 3

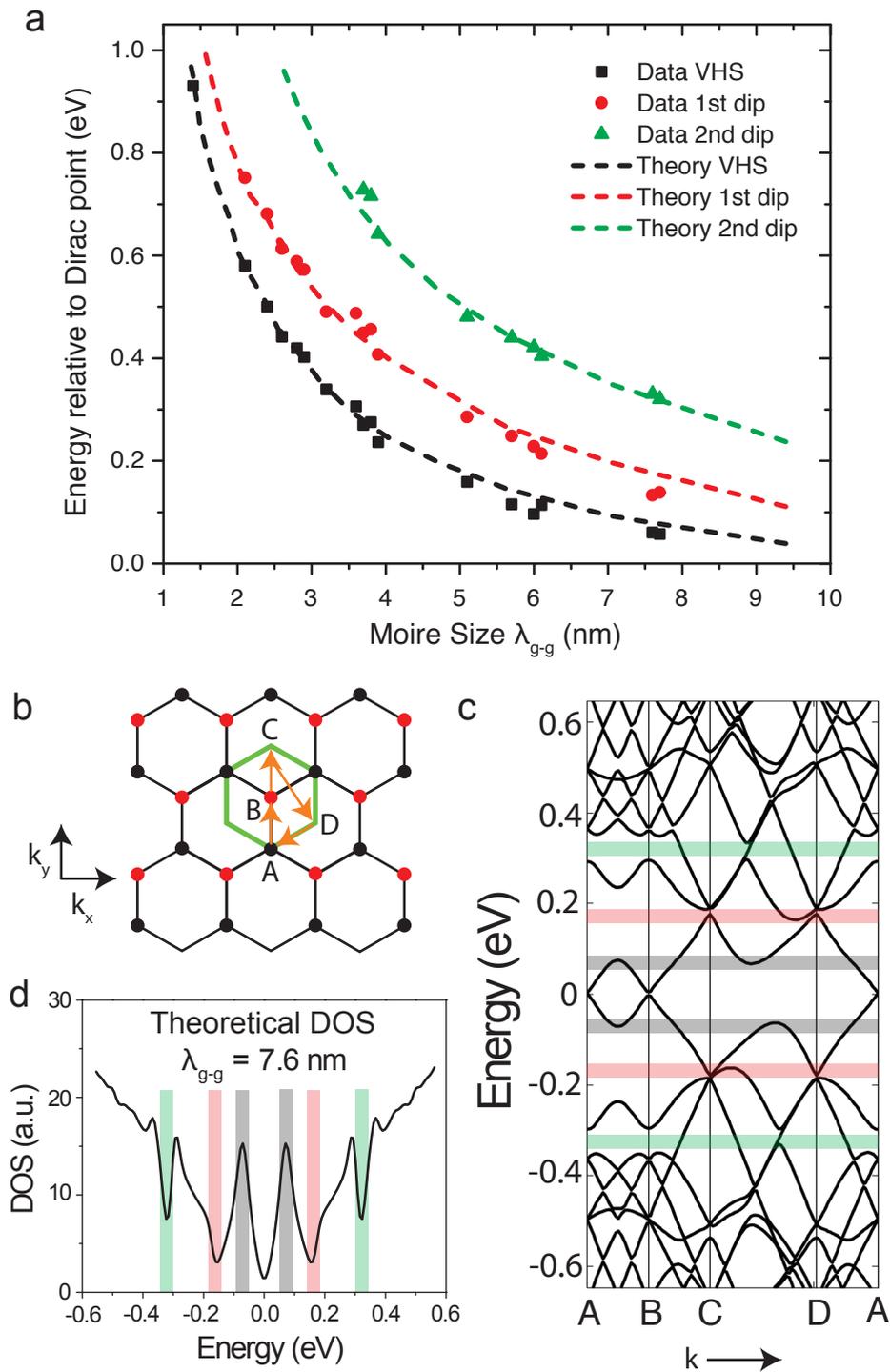

Figure 4